\documentclass[aps,prb,twocolumn,longbibliography,showpacs,superscriptaddress]{revtex4-1}

\usepackage[T1]{fontenc}
\usepackage{soul}
\usepackage{amssymb}
\usepackage{amsmath}  
\usepackage[]{hyperref}

\usepackage{latexsym}
\usepackage{graphicx} 
\usepackage{epstopdf}
\usepackage{comment}
\usepackage{bbold}

\usepackage[toc,page]{appendix}
\usepackage{physics} 
\usepackage{simplewick}
\usepackage{cases}

\begin{document}

\title{Coexistent spin-triplet superconducting and ferromagnetic phases induced by the Hund's rule coupling and electronic correlations II: Effect of applied magnetic field}

\author{M. Fidrysiak}
 \email{maciej.fidrysiak@uj.edu.pl}
\affiliation{Marian Smoluchowski Institute of Physics, Jagiellonian University, ul. {\L}ojasiewicza 11, 30-348 Krak{\'o}w, Poland }
\author{D. Goc-Jag{\l}o}
\email{danuta.goc-jaglo@uj.edu.pl}
 \affiliation{Marian Smoluchowski Institute of Physics, Jagiellonian University, ul. {\L}ojasiewicza 11, 30-348 Krak{\'o}w, Poland }
 \author{E. K\k{a}dzielawa-Major}
 \email{ewa.kadzielawa@doctoral.uj.edu.pl}
 \affiliation{Marian Smoluchowski Institute of Physics, Jagiellonian University, ul. {\L}ojasiewicza 11, 30-348 Krak{\'o}w, Poland }
 \author{P. Kubiczek}
 \email{patryk.kubiczek@physik.uni-hamburg.de}
\affiliation{I. Institut f{\"u}r Theoretische Physik, Universit{\"a}t Hamburg, Jungiusstra{\ss}e 9, D-20355 Hamburg, Germany}
 \author{J. Spa{\l}ek}%
 \email{jozef.spalek@uj.edu.pl}
\affiliation{Marian Smoluchowski Institute of Physics, Jagiellonian University, ul. {\L}ojasiewicza 11, 30-348 Krak{\'o}w, Poland }

\begin{abstract}
  Recently proposed local-correlation-driven pairing mechanism, describing ferromagnetic phases (FM1 and FM2) coexisting with spin-triplet superconductivity (SC) within a single orbitally degenerate Anderson lattice model, is extended to the situation with applied Zeeman field. The model provides and rationalizes in a semiquantitative manner the principal features of the phase diagram observed for $\mathrm{UGe_2}$ in the field absence  [\textit{cf.} Phys. Rev. B \textbf{97}, 224519 (2018)]. As spin-dependent effects play a crucial role for both the ferromagnetic and SC states, the role of the Zeeman field is to single out different stable spin-triplet SC phases. This analysis should thus be helpful in testing the proposed real-space pairing mechanism, which may be regarded as complementary to spin-fluctuation theory suitable for $\mathrm{^3He}$. Specifically, we demonstrate that the presence of the two distinct phases, FM1 and FM2, and associated field-driven \textit{metamagnetic} transition between them, induce \textit{respective metasuperconducting} phase transformation. At the end, we discuss briefly how the spin fluctuations might be incorporated as a next step into the considered here \textit{renormalized} quasiparticle picture. 
\end{abstract}

\maketitle

\section{Introduction}
\label{sec:introduction}
The discovery of the spin-triplet superconductivity (SC) \textit{inside} ferromagnetic (FM) phases of uranium compounds $\mathrm{UGe_{2}}$,\cite{SaxenaNature2000,TateiwaJPhysCondensMatter2001,PfleidererPhysRevLett2002,HuxleyPhysRevB2001} URhGe,\cite{AokiNature2001} UCoGe, \cite{HuyPhysRevLett2007} and UIr\cite{KobayashiPhysicaB2006} is directly related to the question of pairing mechanism and the order-parameter symmetry under such circumstances. Due to substantial correlations in the $f$-electron sector, the situation  here differs from that for superfluid $\mathrm{^3}$He, where a normal (paramagnetic) Landau-Fermi liquid is unstable against the formation of a pure spin-triplet paired state induced by quantum spin fluctuations below the (FM) Stoner instability.\cite{Anderson1978,Vollhardt1990,WysokinskiJPhysCondensMatter2014} The uranium compounds may be regarded as those among the first solid state systems with clear spin-triplet pairing, as the strong effective molecular field acting on spin degrees of freedom in FM phase, at least for $\mathrm{UGe_{2}}$, rules out any spin-singlet SC. Therefore, it is important to see if different phases ($A$, $A_1$,  $A_2$, and $B$) may still appear in an applied magnetic field, in direct correspondence to those observed in ${}^3\mathrm{He}$. Yet, the SC states in the present situation are intertwined with two FM states, FM1 and FM2,\cite{AokiJPhysSocJapan2019} so we would like to single out the different coexisting phases. In brief, the pairing mechanism and order-parameter symmetry for uranium superconductors are yet to be determined in joint theoretical and experimental effort. Here we explicitly identify the possible SC states within the FM and paramagnetic (PM) phases, as well as estimate their gap relative magnitudes.

Recently, we have proposed that pairing in $\mathrm{UGe_{2}}$ emerges due to the combined effect of FM exchange interaction (the Hund's-rule coupling) combined with interelectronic correlations.\cite{Kadzielawa-MajorPhysRevB2018} Ref.~\citenum{Kadzielawa-MajorPhysRevB2018} is regarded as Part~I of our analysis of $\mathrm{UGe_2}$ properties (hereinafter referred to as I). The spin-paired ${A_1}$ state proved to be the dominant phase there with the pair spins opposite to those of average spin polarization, a natural feature appearing in the half-metallic phase.\cite{HaradaPhysRevB2007} Remarkably, within the approach, the ${A_1}$-type SC emerges in a discontinuous manner at the metamagnetic transition between the two distinct FM phases (FM2 $\rightarrow$ FM1), as is evidenced in the recent specific-heat measurements.\cite{TateiwaPhysRevB2004} Finally, SC practically disappears at the boundary of PM phase, which requires invoking a strongly anisotropic and pressure-dependent form of spin-fluctuation spectrum to explain the character of SC state in terms of pairing by long-wavelength FM excitations.\cite{SandemanPhysRevLett2003}
Within our  combined correlation- and exchange-driven pairing scheme all the above features are explained in a unified manner, as both the ferromagnetism and  pairing are directly connected and  driven by the real-space correlations of the same origin.  The changes of applied pressure are theoretically mimicked by us by varying the hybridization magnitude between the almost localized U $5f$ electrons and conduction bands, and regarded as the primary factor inducing the observed evolution.\cite{WysokinskiPhysRevB2014,WysokinskiPhysRevB2015,WysokinskiPhysRevB2016,AbramJMagnMagnMat2016}

Studies of the ground state properties as a function of pressure alone are, however, insufficient to confirm fully the relevance of real-space correlation-driven pairing. This is due to the availability of extensive experimental data covering SC and magnetic properties of $\mathrm{UGe_2}$ in the three-dimensional parameter space spanned by pressure, temperature, and applied magnetic field.\cite{AokiJPhysSocJapan2019} In particular, any proposed pairing mechanism should be minimally tested against the sequence  of magnetic-field-induced simultaneous \textit{metamagnetic} and induced \textit{metasuperconducting} transitions along the first-order line on the field-pressure plane. In this paper we carry out this program and investigate possible \textit{spatially homogeneous} phases in the Zeeman magnetic field. The resultant phase diagram agrees well with available data close to the pressure-induced magnetic transitions. We also provide model band structure in the correlated state, as well as other characteristics, such as the $f$-level filling. The latter parameter points towards almost localized nature of two  out of three $\mathrm{U^{3+}}$ $5f$ electrons and one itinerant, originating from the suggested by us orbital-selective $5f^{3}\rightarrow 5f^{2}$ ($\mathrm{U}^{3+}\rightarrow \mathrm{U}^{4+}$) valence transition. The $f$-state filling is close to an integer, hence the term \textit{almost localized f electrons}. As a reference point, we provide the ground-state results within a more general variational scheme\cite{KubiczekMasterThesis} in zero applied magnetic field and discuss its subsequent simplification (cf. Appendix~\ref{appendix:SGA}). At the end, we outline possible extensions of our approach to incorporate both the full Gutzwiller-type projection (cf. Appendix~\ref{appendix:SGA}) and inclusion of the long-wavelength quantum spin fluctuations (cf. Appendix~\ref{appendix:fluctuations}).

\section{Model and method}
\label{sec:model}

We start from the four-orbital degenerate Anderson lattice model, formulated in the real-space language, that takes the form  

\begin{align}
  \label{eq:hamiltonian}
  \mathcal{H} - &\mu \hat{N}_e = \sum_{i j l \sigma} t_{ij} \hat{c}^{(l)\dagger}_{i\sigma} \hat{c}^{(l)}_{j\sigma} + V \sum_{i l \sigma} \left( \hat{f}^{(l)\dagger}_{i\sigma} \hat{c}^{(l)}_{i\sigma} + \mathrm{H.c.}\right)  \nonumber\\ & + \epsilon^f \sum_{i l} \hat{n}^{f(l)}_i + U \sum_{il} \hat{n}^{f(l)}_{i\uparrow} \hat{n}^{f(l)}_{i\downarrow} + U' \sum_{i} \hat{n}^{f(1)}_i \hat{n}^{f(2)}_i  \nonumber \\ & - 2 J \sum_i \left( \mathbf{\hat{S}}_i^{f(1)} \cdot \mathbf{\hat{S}}_i^{f(2)} + \frac{1}{4} \hat{n}_i^{f(1)} \hat{n}_i^{f(2)} \right) - \mu \hat{N}_e,
\end{align}
where  $\mu$ is the chemical potential for $N_e$-electron $N$-site system, $\hat{f}^{(l)\dagger}_{i\sigma}\; (\hat{f}^{(l)}_{i\sigma})$ is the creation (annihilation) operator of $f$ electron on orbital with $l=1,2$ on site $i$ and spin $\sigma = \uparrow,\downarrow$, hybridized with two species of conduction electrons  characterized by the corresponding operators $\hat{c}^{(l)\dagger}_{i\sigma}$ and $\hat{c}^{(l)}_{i\sigma}$. Additionally, $\hat{n}^{f(l)}_{i\sigma}\equiv \hat{f}^{(l)\dagger}_{i\sigma}\hat{f}^{(l)}_{i\sigma}$ is the particle number operator for $f$ electrons in the original local state $|il\sigma\rangle$ and ${\bf \hat{S}}^{f(l)}_{i}\equiv \left(\hat{S}^{f(l)+}_{i}, \hat{S}^{f(l)-}_{i}, \hat{S}^{f(l)z}_{i}\right)$ denotes the spin operator of $f$ electron on orbital $|il\rangle$. In this minimal model the first term represents $c$-electron hopping, the second an intraatomic hybridization between the subsystems of $f$ and $c$ states, the third is the starting bare atomic $f$-level energy relative to the center of the conduction band. The next two terms express, respectively, the intraorbital and interorbital Coulomb interactions (\textit{both of intraatomic nature}), whereas the third represents ferromagnetic (Hund's-rule) exchange interaction between $f$ electrons. This model has been used by us before to explain the magnetic properties, including classical and quantum criticalities, as well as zero-field SC properties of $\mathrm{UGe_2}$.\cite{Kadzielawa-MajorPhysRevB2018,WysokinskiPhysRevB2015,AbramJMagnMagnMat2016,KubiczekMasterThesis} Here we extend this approach with a detailed analysis of coexisting magnetic and SC properties in  applied Zeemann magnetic field, as well as determine the phase boundaries between them. Note that in applied field two terms should be added to Eq.~\eqref{eq:hamiltonian}: $-g_{f}\mu_0\mu_{B}H\sum_{i}S^{z}_{i}$ and $- g_{c}\mu_0\mu_{B}H\sum_{i}s^{z}_{i}$ for $f$ and $c$ electrons, respectively, where $g_f$ and $g_c$ are gyromagnetic factors, $\mu_0$ denotes material permeability, and $s^{z}_{i}$ is the $z$-th spin component for $c$ electron. Hereafter, for simplicity, we take $g_f = g_c \equiv g$ and introduce reduced field $h \equiv g\mu_0\mu_BH/2$. Moreover, we include only nearest- and next-nearest neighbor hoppings $t < 0$ and $t' = 0.25 |t|$, respectively,  and set $U' = U - 2J$. The total electron filling is taken as $n^\mathrm{tot} = 3.25$. Such a choice yields, at zero field, the sequence of magnetic and SC states that match the experimental phase diagram of $\mathrm{UGe_2}$.\cite{Kadzielawa-MajorPhysRevB2018}

The model \eqref{eq:hamiltonian} is solved within the statistically-consistent Gutzwiller approximation (SGA)\cite{Kadzielawa-MajorPhysRevB2018,WysokinskiPhysRevB2015,AbramJMagnMagnMat2016,KubiczekMasterThesis} that, at zero-temperature, is equivalent to approximate (see below) minimization of the ground-state-energy functional of the form

\begin{align}
E_G \equiv \frac{\langle \Psi_G |\mathcal{H} |\Psi_G \rangle}{\langle \Psi_G |\Psi_G \rangle} \equiv \frac{\langle \Psi_0 |\hat{P}_G\mathcal{H} \hat{P}_G |\Psi_0 \rangle}{\langle \Psi_0 |\hat{P}_G^2|\Psi_0 \rangle}, \label{eq:EG}
\end{align}

\noindent
where the correlated wave function is taken in the form

\begin{align}
  |\Psi_G\rangle \equiv \hat{P}_G |\Psi_0\rangle \equiv \prod_{i\alpha} \hat{P}_{Gi\alpha} |\Psi_0\rangle.
\end{align}

\noindent
$|\Psi_0\rangle$ represents an uncorrelated state, and  $\hat{P}_{Gi\alpha}$ are operators acting locally on orbitals $\alpha \in \{f^{(1)}, f^{(2)}, c^{(1)}, c^{(2)}\}$ at site $i$, introduced to account for local correlations. We adopt a diagonal correlator form $\hat{P}_{Gi\alpha} \equiv   \lambda_{0_{i\alpha}} |{0}_{i\alpha}\rangle\langle {0}_{i\alpha}| + \lambda_{\uparrow_{i\alpha}} |{\uparrow}_{i\alpha}\rangle\langle {\uparrow}_{i\alpha}| + \lambda_{\downarrow_{i\alpha}} |{\downarrow}_{i\alpha}\rangle\langle {\downarrow}_{i\alpha}| + \lambda_{\uparrow\downarrow_{i\alpha}} |{\uparrow\downarrow}_{i\alpha}\rangle\langle{\uparrow\downarrow}_{i\alpha}|$, where the $\lambda$-coefficients serve as variational weights of local many-particle states. Note that $\hat{P}_{Gi\alpha}$ can be generalized to incorporate intraorbital $s$-wave superconducting correlations.\cite{KaczmarczykPhysStatSolidi2015} In our case, however, dominant pairing takes place between distinct $5f$ orbitals, thus we retain the diagonal correlator structure. Moreover, we assume that $\hat{P}_{Gi\alpha}$ respects lattice translational invariance and omit the position index, $i$.

Computation of the expectation values, defined by Eq.~\eqref{eq:EG}, is a complex many-body problem and may be carried out for finite systems by variational Monte-Carlo methods (see., e.g., [\!\!\citenum{BiborskiPhysRevB2018}]) or in thermodynamic limit by suitable diagrammatic-expansion.\cite{BuenemannEurophysLett2012,KaczmarczykNewJPhys2014,FidrysiakJPhysCondensMatter2018} Within the latter framework, eliminating Hartree-bubbles improves substantially series convergence and is achieved by imposing an additional constraint\cite{BuenemannEurophysLett2012,GebhardPhysRevB1990} $\hat{P}_{Gi\alpha}^2 \equiv 1 + x^{(\alpha)} \hat{d}_\mathrm{HF}^{(\alpha)}$, with $\hat{d}_\mathrm{HF}^{(\alpha)} = (\hat{n}^{(\alpha)}_{i\uparrow} - n^{(\alpha)}_{i\uparrow})(\hat{n}^{(\alpha)}_{i\downarrow} - n^{(\alpha)}_{i\downarrow})$, so that only one variational parameter ($x^{(\alpha)}$) prevails per orbital. We used the compact notation $n^{(\alpha)}_{i\sigma} \equiv \langle\hat{n}^{(\alpha)}_{i\sigma}\rangle_0 \equiv \langle\Psi_0|\hat{n}^{(\alpha)}_{i\sigma}|\Psi_0\rangle$. In the above formulation, already the leading diagrammatic contribution tends to capture essential features of correlated lattice models and, to reduce computational cost, we discard higher-order terms (SGA approximation). Moreover, since the correlations are most prominent in the $f$-electron sector, we take $x^{(f^{(1)})} = x^{(f^{(2)})} \equiv x$ and $x^{(c^{(1)})} = x^{(c^{(2)})} = 0$. In the following we skip the orbital indices for the $\lambda$-coefficients as they refer now exclusively to two equivalent $f$ orbitals. For a more complete discussion of the methodological aspects see Appendix~\ref{appendix:SGA}.

One feature of the approach should be underlined at this point. Namely, the average in Eq.~\eqref{eq:EG} involves uncorrelated wave function (in the form of Slater determinant in either direct\cite{ZegrodnikNewJPhys2014} or reciprocal\cite{FidrysiakJPhysCondensMatter2018} space). Therefore, by application of the Wick theorem, the nontrivial averages may be expressed in terms of $n_{i\sigma}^{(\alpha)}$, $\langle\hat{S}^{zf(\alpha)}_i\rangle_0$, $\langle\hat{c}^{(l)\dagger}_{i\sigma} \hat{c}^{(l)}_{j\sigma}\rangle_0$, etc. When executing this procedure, the projection part $\prod_{\alpha, l\neq i, j} \hat{P}_{Gl\alpha}$, acting on the sites which differs from the two-state term in the starting Hamiltonian and generating higher-loop graphs, can be neglected. In effect, we obtain renormalized energy functional
\begin{align}
  \label{eq:energy_functional}
  E_G =   &\sum_{i j l \sigma} t_{ij} \langle\hat{c}^{(l)\dagger}_{i\sigma} \hat{c}^{(l)}_{j\sigma}\rangle_0 + V \sum_{i l \sigma} q_{\sigma} \left( \langle\hat{f}^{(l)\dagger}_{i\sigma} \hat{c}^{(l)}_{i\sigma}\rangle_0 + \mathrm{c.c.}\right) \nonumber \\ + & \sum_{i \sigma}\left[U'  g_{1\sigma} + (U' - J)  g_{2\sigma}\right] |\langle\hat{f}^{(1)}_{i\sigma} \hat{f}^{(2)}_{i\sigma}\rangle_0|^2  \nonumber \\ - & \sum_i 2J \langle \hat{S}^{z f(1)}_i \rangle_0 \langle \hat{S}^{z f(2)}_i \rangle_0  \nonumber \\ + & \sum_i (U' - \frac{J}{2}) \langle\hat{n}^{f(1)}_i\rangle_0 \langle\hat{n}^{f(2)}_i\rangle_0   \nonumber\\ & +  \sum_{i l \sigma} (\epsilon^f - h\sigma) \langle\hat{n}^{f(l)}_{i\sigma}\rangle_0 + U  \sum_{il} \lambda_{\uparrow \downarrow}^2 \langle\hat{n}^{f(l)}_{i\uparrow}\rangle_0 \langle\hat{n}^{f(l)}_{i\downarrow}\rangle_0 \nonumber \\ & - h \sum_{il\sigma} \sigma \langle\hat{n}^{c(l)}_{i\sigma}\rangle_0, 
\end{align}

\noindent
where the renormalization factors

\begin{align}
  \label{eq:renormalization_factors}
g_{1\sigma} \equiv & 2  (\lambda_{\uparrow\downarrow}^2 - \lambda_{\bar{\sigma}}^2)(\lambda_\sigma^2 + (\lambda_{\uparrow\downarrow}^2 - \lambda_\sigma^2)  n^{f(l)}_{\bar{\sigma}}) n^{f(l)}_{\bar{\sigma}}, \nonumber \\  g_{2\sigma} \equiv & (\lambda_{\uparrow\downarrow}^2 - \lambda_{\bar{\sigma}}^2)^2  \left(n^{f(l)}_{\bar{\sigma}}\right)^2 + \left(\lambda_\sigma^2 + (\lambda_{\uparrow\downarrow}^2 -
                                                                                                                                                                                                                                                 \lambda_\sigma^2)  n^{f(l)}_{\bar{\sigma}}\right)^2, \nonumber
\end{align}
\noindent
and
\begin{align}                                                                                                                              q_\sigma \equiv& \lambda_0 \lambda_\sigma + (\lambda_{\uparrow\downarrow} \lambda_{\bar{\sigma}} - \lambda_0 \lambda_\sigma)  n^{f(l)}_{\bar{\sigma}}
\end{align}

\noindent
appear in response to local electronic correlations ($\bar{\sigma} \equiv -\sigma$).\cite{Kadzielawa-MajorPhysRevB2018}

This renormalized Hamiltonian of the single-quasiparticle type with pairing should thus be diagonalized first, before the ground state energy \eqref{eq:EG} in the correlated state is evaluated explicitly. Equivalently, optimization of $E_G$ over wave function $|\Psi_0\rangle$ yields an effective non-linear Schr\"{o}dinger equation $\mathcal{H}_\mathrm{eff}|\Psi_0\rangle = E|\Psi_0\rangle$ with the following effective Hamiltonian

\begin{align}
	\mathcal{H}_\mathrm{eff}  = \sum_{\mathbf{k}, \sigma}^{} \Psi_{\mathbf{k} \sigma}^\dagger \left( \begin{array}{cccc}
		\epsilon_{ \mathbf{k} \sigma } & 0 &  q_\sigma V&  0\\
		0 & - \epsilon_{ \mathbf{k} \sigma } & 0 & - q_\sigma V\\
		q_\sigma V & 0 & \epsilon_{ \sigma }^{f} & \Delta_{ \sigma \sigma }^{ff} \\
		0 & - q_\sigma V & \Delta_{ \sigma \sigma }^{ff} & -\epsilon_{ \sigma }^{f} \\
	\end{array}
	\right) \Psi_{\mathbf{k} \sigma}
  + E_0,
  \label{eq:effective_hamiltonian}
\end{align}

\noindent
which is expressed now in terms of Nambu spinor $\Psi_{\mathbf{k}\sigma }^\dagger \equiv \left( \hat{c}^{(1)\dagger}_{ \mathbf{k} \sigma }, \hat{c}^{(2)}_{ -\mathbf{k} \sigma}, \hat{f}^{(1)\dagger}_{ \mathbf{k} \sigma }, \hat{f}^{(2)}_{ -\mathbf{k} \sigma } \right)$ and leads to the expectation value \eqref{eq:EG}. Here

\begin{align}
\epsilon_{ \mathbf{k}\sigma } =&  2 t [\cos(k_x) + \cos(k_y)] \nonumber \\ &+ 4t^\prime \cos(k_x)\cos(k_y) - \mu - h\sigma
  \label{eq:zeeman_split_dispersion}
\end{align}

\noindent
is the Zeeman-split tight-binding dispersion relation for bare conduction electrons, 

        	\begin{align}
		\epsilon_{ \sigma }^f &\equiv \frac{\partial E_G}{\partial n^{f(1)}_{i \sigma}}  =                  
		\epsilon^f + U \lambda_{\uparrow \downarrow}^2  n^{f(1)}_{i \bar{\sigma}}+ ( U^\prime - J ) n^{f(2)}_{i \sigma} + U^\prime n^{f(2)}_{i \bar{\sigma}}  \nonumber \\&
		+ 
		\left( \frac{\partial q_{\bar{\sigma}}}{\partial n^{f(1)}_{i \sigma}} V \sum_l\langle\hat{f}^{(l)\dagger}_{i\bar{\sigma}} \hat{c}^{(l)}_{i\bar{\sigma}}\rangle_0 + \mathrm{c.c.} \right)	
		 \nonumber \\ &+
		\left( \frac{\partial g_{1\bar{\sigma}}}{\partial n^{f(1)}_{i \sigma}} U^\prime + \frac{\partial g_{2\bar{\sigma}}}{\partial n^{f(1)}_{i \sigma}} (U^\prime - J) \right) |\langle\hat{f}^{(1)}_{i\bar{\sigma}} \hat{f}^{(2)}_{i\bar{\sigma}}\rangle_0|^2 \nonumber\\&- \mu - h\sigma
	\end{align}

\noindent
determines position of the renormalized $f$-electron level, and $E_0$ is the energy shift (that does not influence expectation values but contributes to the ground state energy).  The effective gap parameter, $\Delta_{ \sigma \sigma }^{ff}$, and the effective SC coupling constant $\mathcal{V}_{\sigma}$ (to be addressed below),  are defined by the relation

        \begin{align}
          \label{eq:delta_sigma_sigma}
\Delta_{ \sigma \sigma }^{ff} \equiv& \mathcal{V}_{\sigma}   \langle \hat{f}^{(1)}_{i\sigma} \hat{f}^{(2)}_{i\sigma}\rangle_0 \equiv \frac{\partial E_G}{\partial \langle\hat{f}^{(1)\dagger}_{i\sigma} \hat{f}^{(2)\dagger}_{i\sigma}\rangle_0}  = \nonumber \\ & = - \left[g_{1\sigma} U^\prime + g_{2\sigma} (U^\prime - J)\right]   \langle\hat{f}^{(1)}_{i\sigma} \hat{f}^{(2)}_{i\sigma}\rangle_0.
        \end{align}

\noindent
The  resultant integral Schr\"{o}dinger-type equation is solved numerically in the loop with minimization of the energy functional [Eq.~\eqref{eq:energy_functional}] over the correlator parameter $x$. In order to avoid finite-size effects that become severe for weak SC order, considered here, we performed the calculations directly in the thermodynamic limit using adaptive integration. Note that the effective pairing potential $\mathcal{V}_\sigma$ can be attractive even in the regime where its Hartree-Fock (unrenormalized) correspondant $\mathcal{V}^\mathrm{HF}_\sigma = U-3J$ is already repulsive.\cite{Kadzielawa-MajorPhysRevB2018} In that case, the pairing is induced by nontrivial correlation effects.

A methodological remark is in place at this point. The above scheme employs correlator $\hat{P}_G$ that acts \textit{separately} on each orbital. In a multi-band system, such as the one considered here, one could expect that the correlator should allow for more general many-body states involving multiple orbitals at once. Such an extension makes it difficult to achieve numerical accuracy required to study SC order emerging on the scale of the order of one kelvin in uranium materials. We have, nonetheless, performed such an extended analysis\cite{KubiczekMasterThesis} for zero field and limited range of model parameters, with the results very close to those obtained from the above simplified scheme. The discussion of those formal issues is deferred to Appendix~\ref{appendix:SGA}.

\begin{figure}
  \includegraphics[width = 1.0\columnwidth]{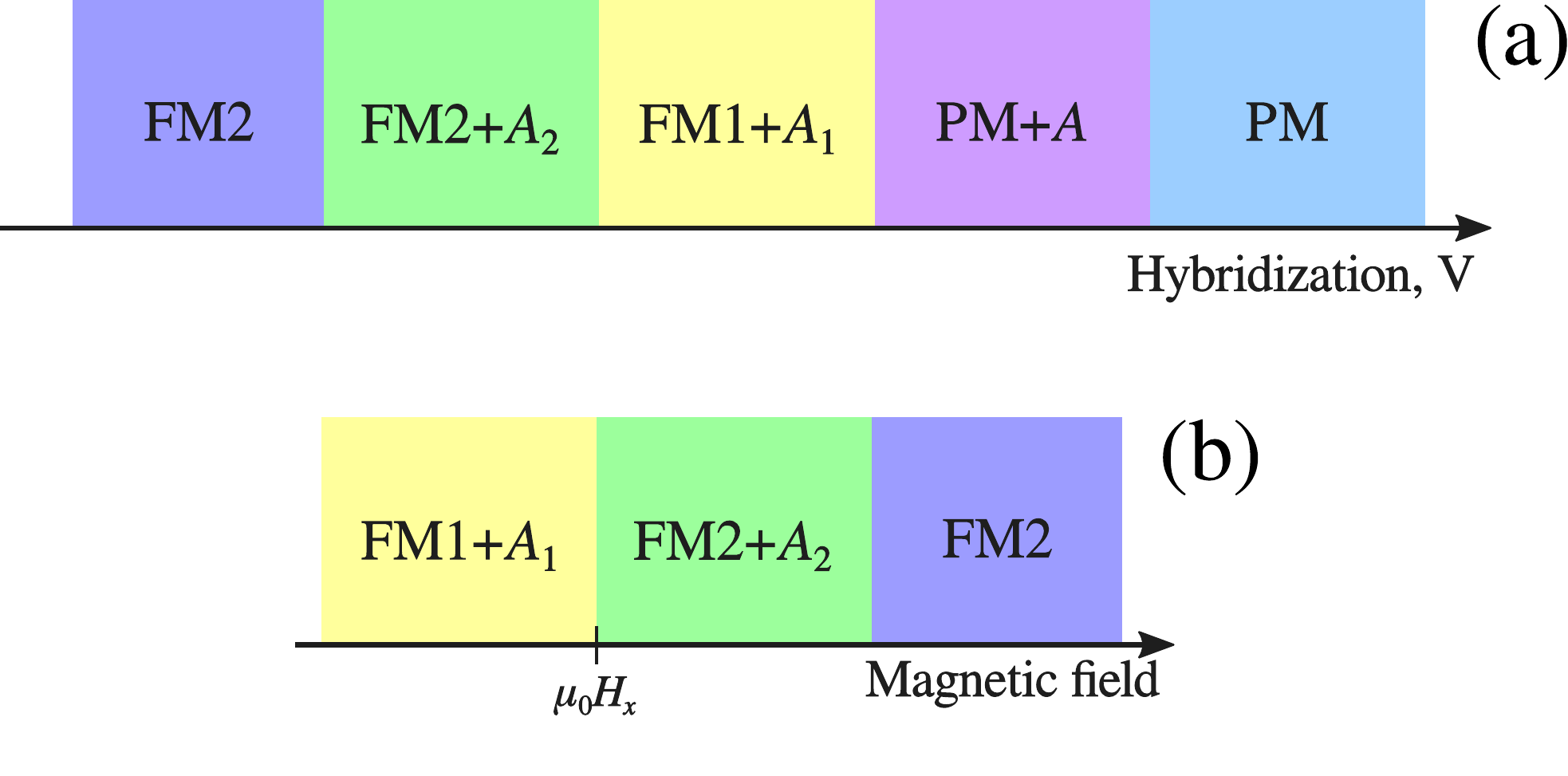}
  \caption{(a) Sequence of ferromagnetic (FM2, FM1) and nonmagnetic (PM) phases, coexisting with  superconducting (${A_2}$, ${A_1}$, $A$) states, as a function of increasing hybridization magnitude (emulating pressure change\cite{AbramJMagnMagnMat2016}) for zero external magnetic field. (b) The same as in (a), but for fixed hybridization and varying pressure, with the most prominent FM1+$A_1$ phase taken as the starting point. The boundaries mark transition points in both the magnetic and  superconducting sectors.}
  \label{fig:phases}
\end{figure}

\begin{table}
  \centering
  \caption{Detailed structure of the anomalous local $f$-$f$ amplitudes for various coexistent magnetic and superconducting phases, appearing for the four-orbital periodic Anderson model~\eqref{eq:hamiltonian}.}
  \begin{tabular}{cc}
    Phase & Anomalous $f$-$f$ amplitudes \\
    \hline\hline
    $\mathrm{PM}+A$ & $\langle f_{i\downarrow}^{(1)\dagger} f_{i\downarrow}^{(2)\dagger}\rangle_0 = \langle f_{i\uparrow}^{(1)\dagger} f_{i\uparrow}^{(2)\dagger}\rangle_0 > 0$\\
    $\mathrm{FM1}+A_1$ & $\langle f_{i\downarrow}^{(1)\dagger} f_{i\downarrow}^{(2)\dagger}\rangle_0 > 0$,  $\langle f_{i\uparrow}^{(1)\dagger} f_{i\uparrow}^{(2)\dagger}\rangle_0 = 0$ \\
    $\mathrm{FM2}+A_2$ & $\langle f_{i\downarrow}^{(1)\dagger} f_{i\downarrow}^{(2)\dagger}\rangle_0 > \langle f_{i\uparrow}^{(1)\dagger} f_{i\uparrow}^{(2)\dagger}\rangle_0 > 0$
  \end{tabular}
  \label{tab:sc_amplitude_definitions}
\end{table}

\section{Results and discussion}
\label{sec:results}

\subsection{Zero-field results as a reference point}

The SC pairing discussed here is of local interorbital nature, i.e., of odd parity in the orbital and even in the spin indices, as was proposed before.\cite{SpalekPhysRevLett1987,KubiczekMasterThesis} In Fig.~\ref{fig:phases}(a) we draw schematically the sequence of phases obtained in zero applied field. The three SC states are labeled in a similar manner as those for the case of superfluid $\mathrm{^3He}$, with ${A_1}$ being fully spin polarized ($\downarrow\downarrow$ Cooper pairs only) and ${A_2}$ phase is that with unequal order parameter amplitudes $\uparrow\uparrow$ and $\downarrow\downarrow$ which finally equalize in the PM state and result in $A$-type SC. Formally, the above SC states are characterized by non-vanishing anomalous amplitudes detailed in Table~\ref{tab:sc_amplitude_definitions}. The coexistent phase FM1+${A_1}$ is the most prominent, ${A_2}$ and ${A}$ states play only a very minor, if not negligible role in the field absence. We emulate changing external pressure by the corresponding change in hybridization magnitude (for a detailed discussion of this particular point see Part I and Ref.~[\!\!\!\citenum{AbramJMagnMagnMat2016}]). As shown below, the role of the field is to enhance the presence of the $A_2$ phase.

For the sake of completeness, in Table~\ref{table:gap_values} we provide selected numerical values of the effective SC gap parameters for $H=0$ in the ${A_1}$, ${A_2}$, and $A$ phases.  They are defined as partial derivatives of the variational functional with respect to anomalous amplitudes [cf. Eqs.~\eqref{eq:delta_sigma_sigma}, \eqref{eq:def_Delta_ff}, and \eqref{eq:def_Delta_fc}] and physically determine the spectrum of projected quasiparticle excitations.\cite{FidrysiakJPhysCondensMatter2018} As the SC transition sequence ${A_2}\rightarrow A_1 \rightarrow A$ takes place simultaneously with that corresponding to discontinuous magnetic transitions (FM2$\rightarrow$FM1$\rightarrow$PM), they are also discontinuous, but the former discontinuities are probably too weak to be detected experimentally (note that the maximal value of the SC transition temperature does not exceed $1\,\mathrm{K}$ in all uranium systems).\cite{SaxenaNature2000,TateiwaJPhysCondensMatter2001,PfleidererPhysRevLett2002,HuxleyPhysRevB2001,AokiNature2001,HuyPhysRevLett2007,KobayashiPhysicaB2006} However, we obtain a clear sign of \textit{metasuperconducting} transition accompanying the corresponding metamagnetic jumps. This issue is discussed below.

\setlength{\tabcolsep}{4pt}
\begin{table}
   \caption{Superconducting gap components as a function of hybridization $|V|$ for $U = 4|t|$ and $J = 1.6 |t|$. Estimated numerical accuracy $\delta\Delta^{ff}_{\sigma\sigma}/|t|$ is also provided in the last column.}
     \label{table:gap_values}
    \begin{tabular}{cccc}
  $V/t$ & $100\times\Delta^{ff}_{\downarrow\downarrow}/|t|$ & $100\times\Delta^{ff}_{\uparrow\uparrow}/|t|$ & $100 \times \delta\Delta^{ff}_{\sigma\sigma}/|t|$ \\
  \hline
  \hline
1.1666667 & 0.0000000 & 0.0000000 & 0.0000011\\
1.3000000 & 0.0038378 & 0.0000000 & 0.0000012\\
1.3333333 & 0.0225363 & 0.0000000 & 0.0000012\\
1.4000000 & 0.5660415 & 0.0001295 & 0.0000013\\
1.4500000 & 5.8775861 & 0.0000000 & 0.0000014\\
1.5000000 & 5.1822010 & 0.0000000 & 0.0000014\\
1.5500000 & 4.5934998 & 0.0000000 & 0.0000013\\
1.6000000 & 4.0906100 & 0.0000000 & 0.0000013\\
2.0000000 & 1.8155386 & 0.0000000 & 0.0000012\\
2.5000000 & 0.7775326 & 0.0000000 & 0.0000011\\
3.0000000 & 0.3706006 & 0.0000000 & 0.0000011\\
3.5000000 & 0.1909743 & 0.0000000 & 0.0000010\\
4.0000000 & 0.1049818 & 0.0000000 & 0.0000010\\
4.1500000 & 0.0887245 & 0.0000000 & 0.0000010\\
4.1000000 & 0.9525063 & 0.9525063 & 0.0000011\\
4.2000000 & 0.8016905 & 0.8016935 & 0.0000011\\
4.2500000 & 0.7355641 & 0.7355662 & 0.0000010\\
4.4000000 & 0.5685009 & 0.5685018 & 0.0000010\\
5.0000000 & 0.2062559 & 0.2062561 & 0.0000010\\
\end{tabular}
\label{gaps}
\end{table}

\subsection{Discontinuous phase transition in an applied magnetic field}

\begin{figure}
  \includegraphics[width = .9\columnwidth]{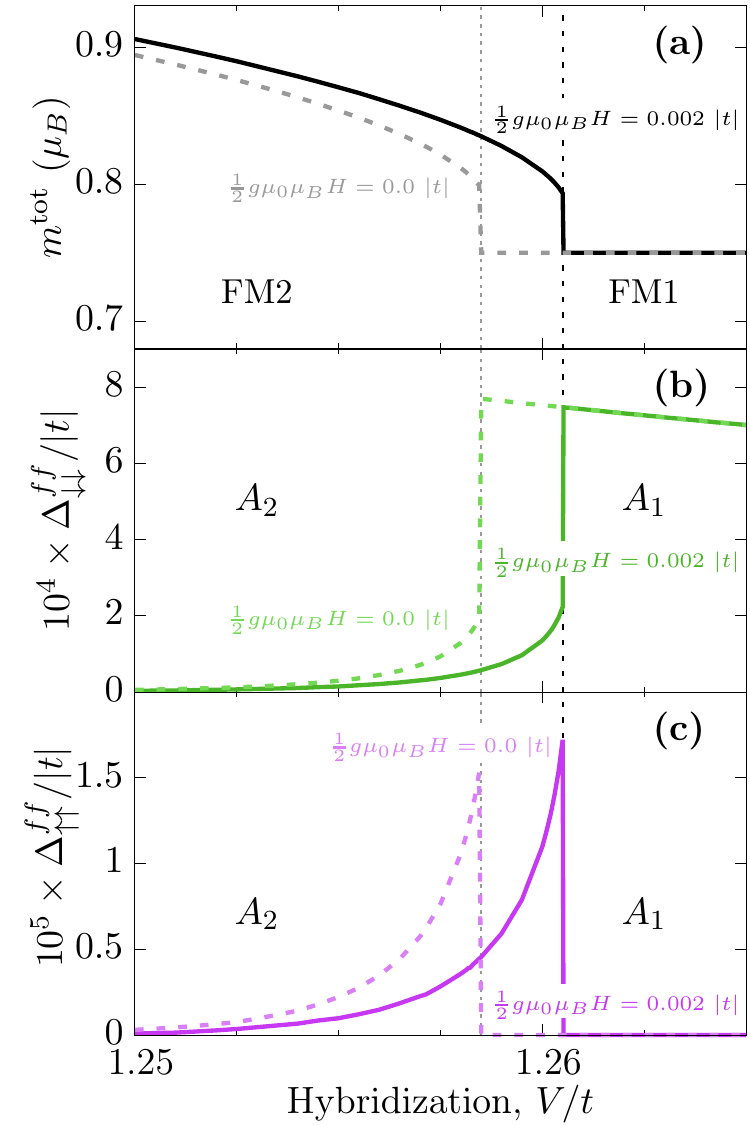}
	\caption{(a) Total magnetization, $m^\mathrm{tot}= m^{f}+m^{c}$ (a) and superconducting gap  components [(b and (c)] vs. hybridization magnitude. Solid lines correspond to field value $h/|t|=0.002$, the dashed lines represent $h=0$ situation.\cite{Kadzielawa-MajorPhysRevB2018} The  microscopic parameters are: $U/|t| = 3.5$, $J/|t| = 1.1$, $T = 0\,\mathrm{K}$ , $t^\prime/|t| = 0.25$, $\epsilon^f/|t| = -4$, $n^{\mathrm{tot}}\equiv n^{f}+n^{c} = 3.25$, and the field $h=0.002|t|$, which corresponds to $\mu_0 H = 6.9\,\mathrm{T}$ for the nearest-neighbor hopping  $|t|=0.2$ eV.}
	\label{fig:phase_diagram_J11_H0002}
\end{figure}

In Fig.~\ref{fig:phases}(b) we have drawn schematically the  sequence of phases appearing with the increasing applied field, starting from the most prominent FM1+${A_1}$. The high-field phase is always pure high-moment FM2. To illustrate the situation quantitatively, we have plotted  in Fig.~\ref{fig:phase_diagram_J11_H0002}(a)-(c) the total magnetic moment $m^\mathrm{tot}\equiv m^{f}+m^{c}$  [cf. panel (a)], the $\Delta^{ff}_{\downarrow\downarrow}$ [panel (b)], and $\Delta^{ff}_{\uparrow\uparrow}$ [panel (c)] SC amplitudes. All the transitions are of discontinuous \textit{metamagnetic/metasuperconducting} character. The paired states disappear gradually as the magnetic moment increases in the FM2 phase. In that final state, both the diagonal pairing correlation $\langle \hat{f}^{(1)\dagger}_{i\sigma} \hat{f}^{(2)\dagger}_{i\sigma}\rangle_0$ and spin fluctuations are suppressed by the magnetic-field-enforced moment saturation. In Fig.~\ref{fig:phase_diagram_J11_H0002} we show a representative situation near the FM2-FM1 boundary. Note that the considered discontinuous transitions may be easier to detect with the help of magnetic methods rather than by specific-heat measurements.

\begin{figure}
  \centering
  \includegraphics[width = .8\columnwidth]{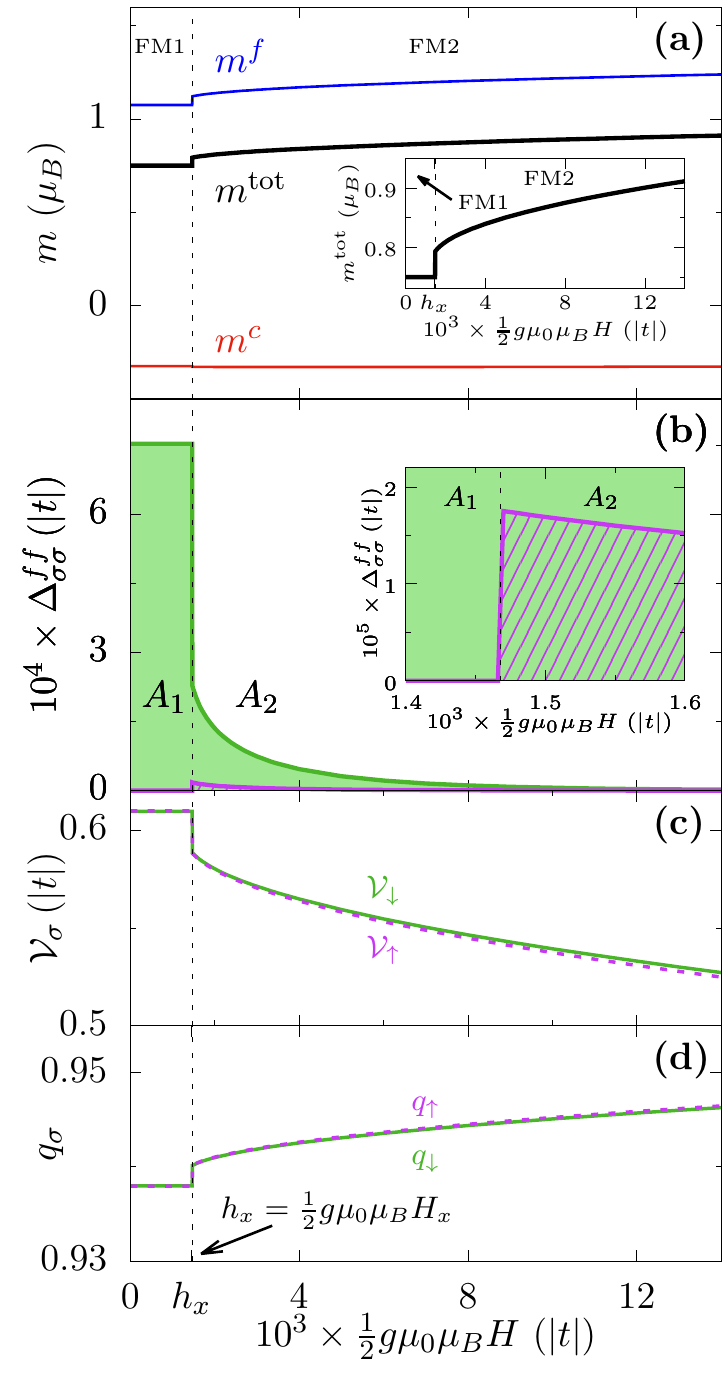}
	\caption{Selected properties in applied magnetic field at zero temperature. (a) Moments: $m^{\mathrm{tot}}$ -- total (black line), $m^f$ -- $f$-electron component (blue), $m^c$ -- $c$-electron component (red). (b) Spin-triplet $f$-$f$ superconducting gap components: $\Delta^{ff}_{\uparrow \uparrow}$ -- purple, $\Delta^{ff}_{\downarrow \downarrow}$ -- green. (c) Spin-dependent pairing potential $\mathcal{V}_\sigma$. (d) Renormalization factors $q_\sigma$ [cf. Eq.~\eqref{eq:renormalization_factors}]. Phase transition from FM1+$A_1$ to FM2+$A_2$ takes place at $h_x/|t| = 0.001468$, which corresponds to magnetic field $\mu_0 H_x \approx 5.1\,\mathrm{T}$ for $|t| = 0.2\,\mathrm{eV}$. The results are obtained for the set of parameters: $U/|t| = 3.5$, $J/|t| = 1.1$, $T = 0\ $K, $V/t=1.26$, $t^\prime/|t| = 0.25$, $\epsilon^f/|t| = -4$, $n^{\mathrm{tot}}=3.25$. Insets in (a) and (b) detail the discontinuous nature of the transitions. The pairing coupling constant is only weakly spin-dependent, whereas the gaps are due to strong spin dependence of the electronic structure (see the text).}
	\label{fig:phase_diagram_J11_V-126}
\end{figure}

\begin{figure}
  \includegraphics[width = 0.9\columnwidth]{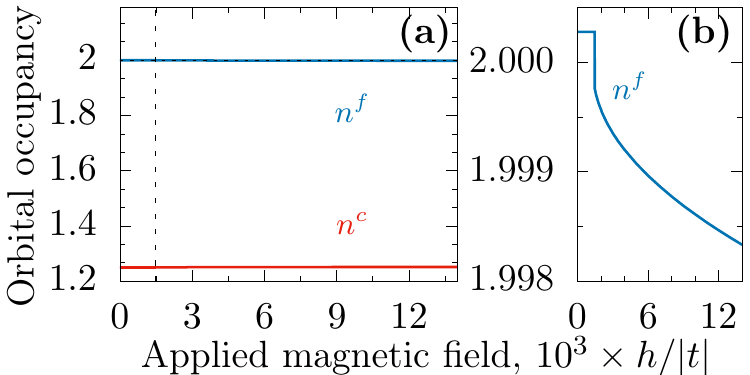}
	\caption{Occupancies $n^f$ and $n^c$ as a function of applied magnetic field for parameters the same as those adopted in Fig.~\ref{fig:phase_diagram_J11_V-126}. The $f$-orbital occupancy is almost equal to unity showing an almost localized nature of those electrons even in the presence of a sizable hybridization.}
	\label{fig:occupancy}
\end{figure}

\begin{figure}
  \includegraphics[width = 0.9\columnwidth]{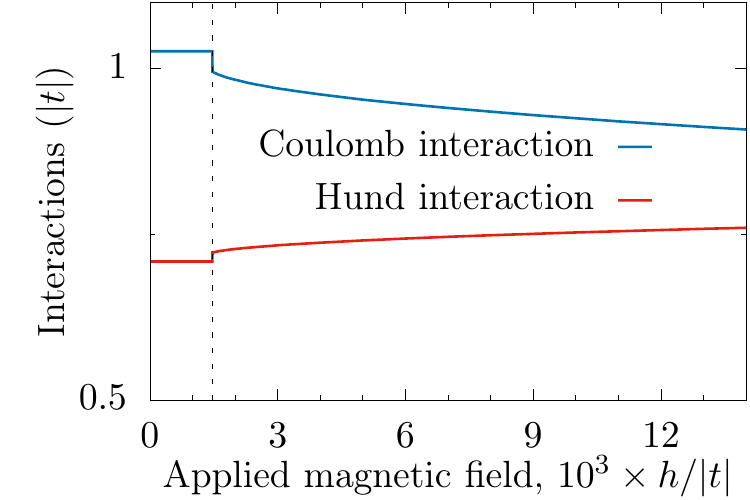}
	\caption{Relative contributions of the renormalized Hund's-rule coupling and direct intraorbital Coulomb interactions to the ground state energy for the value of hybridization $V/t=1.26$, i.e., at the threshold of FM2$\rightarrow$FM1 transition, where the $A_1$ SC phase appears in an abrupt manner. The model parameters coincide with those adopted in Fig.~\ref{fig:phase_diagram_J11_V-126}.} 
	\label{fig:interactions}
\end{figure}

The character of the transitions as a function of applied field for fixed value of hybridization is provided  in Fig.~\ref{fig:phase_diagram_J11_V-126}. In Fig.~\ref{fig:phase_diagram_J11_V-126}(a) the components of the total moment are displayed. Note that the negative  (Kondo-like) $c$-electron magnetic moment is practically field independent. In Fig.~\ref{fig:phase_diagram_J11_V-126}(b) a pronounced ${A_1}\rightarrow{A_2}$ SC transition region is emphasized. Insets to Fig.~\ref{fig:phase_diagram_J11_V-126}(a) and (b) are to visualize clearly the discontinuities. This behavior may be compared with the measurements of the upper critical field $H_{c2}$ as a function of temperature  close to the field-induced metamagnetic FM1$\rightarrow$FM2 transition point (cf. Fig.~10 of Ref.~[\!\!\citenum{HuxleyPhysRevB2001}]). Specifically, for $13.5\,\mathrm{kbar}$ a sharp drop of SC transition temperature is observed experimentally above $\mu_0 H_x \approx 2\,\mathrm{T}$, in qualitative agreement with theoretical result depicted in panel (b). Also, the SC state is detected unambiguously on both sides of the transition as is predicted by the local-correlation pairing scenario elaborated here. In Fig.~\ref{fig:phase_diagram_J11_V-126}(c) we plot the spin-dependent effective coupling constants $\mathcal{V}_{\sigma}$ (cf. I), defined by the relation $\Delta^{ff}_{\sigma\sigma}\equiv \mathcal{V}_{\sigma}   \langle \hat{f}^{(1)}_{i\sigma} \hat{f}^{(2)}_{i\sigma}\rangle_0$ [cf. Eq.~\eqref{eq:delta_sigma_sigma}].
Note that the pairing potential (effective coupling constant) is only moderately spin-dependent in applied field, whereas the spin-components of the gap behave in very different manner.  Such an asymmetry of the results for $\sigma=\uparrow, \downarrow$ components of $\Delta^{ff}_{\sigma\sigma}$ can be easily understood as, e.g.,  in the FM1 phase the spin-majority subband is full and the system becomes half-metallic, which implies $\Delta^{ff}_{\uparrow\uparrow}\equiv 0$. Finally, in panel (d) we plot the factors $q_\sigma$ that renormalize $f$-$c$ hybridization magnitude [cf. Eqs.~\eqref{eq:effective_hamiltonian} and \eqref{eq:renormalization_factors}]. These coefficients turn out to be of the order of unity, hence their major is to renormalize the pairing coupling constant rather than the single-particle dynamics.

\begin{figure}
  \centering
  \includegraphics[width = .9\columnwidth]{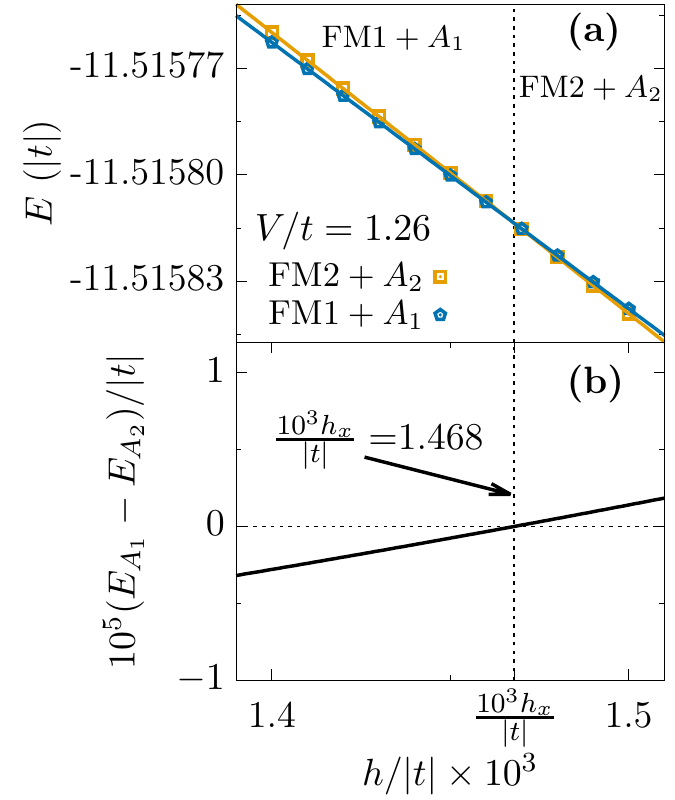}
  \caption{
The ground-state energies of the two phases marked (a) with the phase-transformation point marked by the vertical dashed line. The difference of energies of the two phases (b) is of the order of $0.1\,\mathrm{K}$. Note also that the transition is discontinuous as the two lines in (a) have at $h_x$ slightly different slopes.}
\label{fig:energies}
\end{figure}

\begin{figure}
  \includegraphics[width = 0.7\columnwidth]{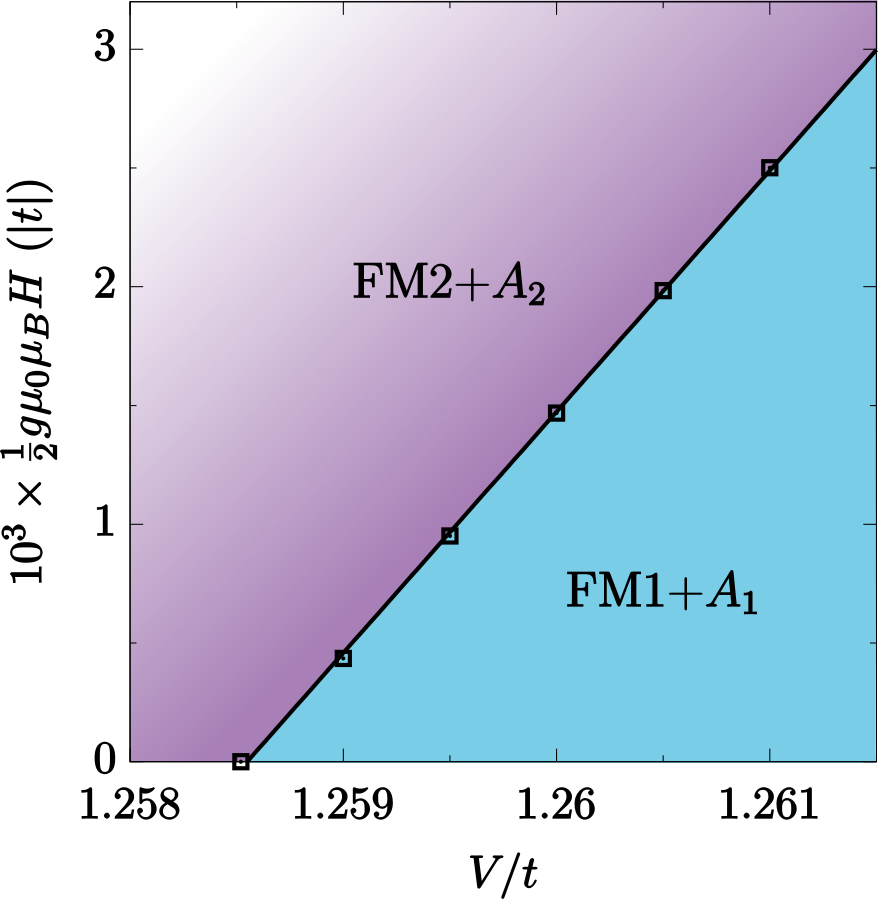}
  \caption{Calculated $f$-$c$ hybridization dependence of the characteristic transition field $\mu_0 H_x$ in the vicinity of the FM2-FM1 metamagnetic instability for $H=0$. The weak discontinuity may be detected by the magnetic susceptibility measurements across the boundary for fixed pressure ($V/t$ ratio).}
  \label{fig:Hx}
\end{figure}

To complete the picture, we have plotted in Fig.~\ref{fig:occupancy}(a)-(b) the overall $f$-level occupancy ($n^f$) and that of $c$ band ($n^c$); the details of $n^f$ evolution are shown in panel~(b). The $f$-orbital occupancy is very close to unity, showing that those electrons have \textit{an almost localized nature}. Moreover, approximately one additional electron (per $\mathrm{U}$) is effectively transferred to the conduction band (strictly speaking $n^c \approx 1.25$). This conclusion confirms again our conjecture reached before \cite{Kadzielawa-MajorPhysRevB2018} that in the case of $\mathrm{U^{3+}}$ each ion effectively turns into into $\mathrm{U^{4+}}$ with two nearly localized electrons and an itinerant electron created at the same time. Also, the Hund's-rule and intraorbital Coulomb-interaction contributions to the total energy depend relatively strongly on the value of applied field, as demonstrated in Fig.~\ref{fig:interactions}. This feature supports further the strongly correlated nature of the system, in which various contributions balance out (partially compensate) each other in such a manner that a much smaller Zeeman contribution plays the role of a tip of balance between the localized and itinerant states of $f$-electrons.\cite{SpalekPhysRevLett1987} Such a circumstance is characteristic of \textit{a Hund metal}, as elaborated in I. Additionally, the SC in the relevant uranium systems emerges at low temperatures, typically  below $1\,\mathrm{K}$.\cite{SaxenaNature2000,TateiwaJPhysCondensMatter2001,PfleidererPhysRevLett2002,HuxleyPhysRevB2001,AokiNature2001,HuyPhysRevLett2007,KobayashiPhysicaB2006} The related discontinuous transformations in the field involve even more subtle free-energy changes, as shown explicitly in Fig.~\ref{fig:energies} for the situation with FM1+${A_1}$ $\rightarrow$ FM2+${A_2}$ phase transformation. The corresponding total-energy change is of the order $10^{-5} |t|$, which for $|t| \sim 0.2$ eV is below the scale of $0.1\,\mathrm{K}$. Nevertheless, the accuracy of our numerical results is well below these values (cf. Fig.~6 of Part I\cite{Kadzielawa-MajorPhysRevB2018}).

Finally, in Fig.~\ref{fig:Hx} we plot the boundary line between the FM1+$A_1$ and FM2+$A_2$ phases in the $H$-$|V|$ plane. It has a linear character in this narrow range of $V/t$ encompassing the FM2-FM1 discontinuous metamagnetic transition at $H = 0$ as a starting point. This borderline may serve as an important feature and, in particular, help to single out the relevant SC phases with symmetry of the order parameter ($A_1 \rightarrow A_2$) changing in a discontinuous manner.

In summary, the observed SC discontinuities in an applied magnetic field are relatively small. However, with the help of sensitive magnetic measurements of ac susceptibility, they should be detectable. Also, the appearance of the second component of the SC gap at the $A_1\rightarrow A_2$ transition may become observable in the pair tunneling spectroscopy. These rather simple remarks  require though a more quantitative  substantiation.

\subsection{Electronic structure}
 
\begin{figure*}
  \includegraphics[width = 1.8\columnwidth]{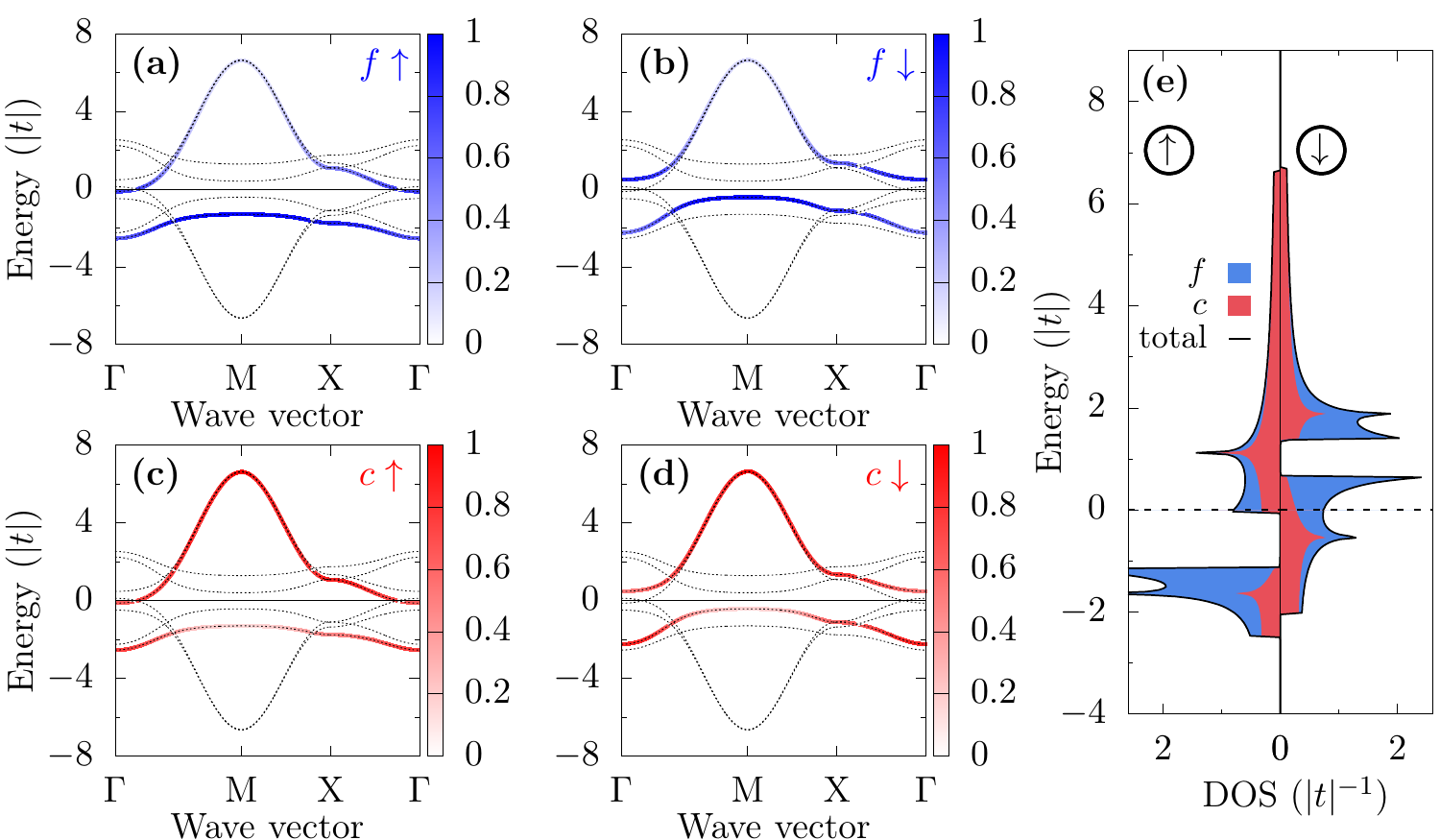}
  \caption{Renormalized band structure for $h/|t| = 0.002$ and $V/t = 1.26$ in the FM2+$A_2$ phase (set of other parameters:  $U/|t| = 3.5$, $J/|t| = 1.1$, $ T = 0\,\mathrm{K}$, $n^{\mathrm{tot}} = 3.25$, $t'/|t| = 0.25 $, $\epsilon^f/|t| = -4$). The eigenenergies are represented by dotted lines. The partial spectral-weight contributions from $f$-electrons [(a) and (b)] are marked in blue, whereas those for $c$-electrons [(c) and (d)] in red. Color intensity represents the spectral-weight magnitude. In panel (e) spin-resolved density of states is presented.}
	\label{fig:band_structure_U35_J11_FM2}
\end{figure*}

\begin{figure*}
  \includegraphics[width = 1.8\columnwidth]{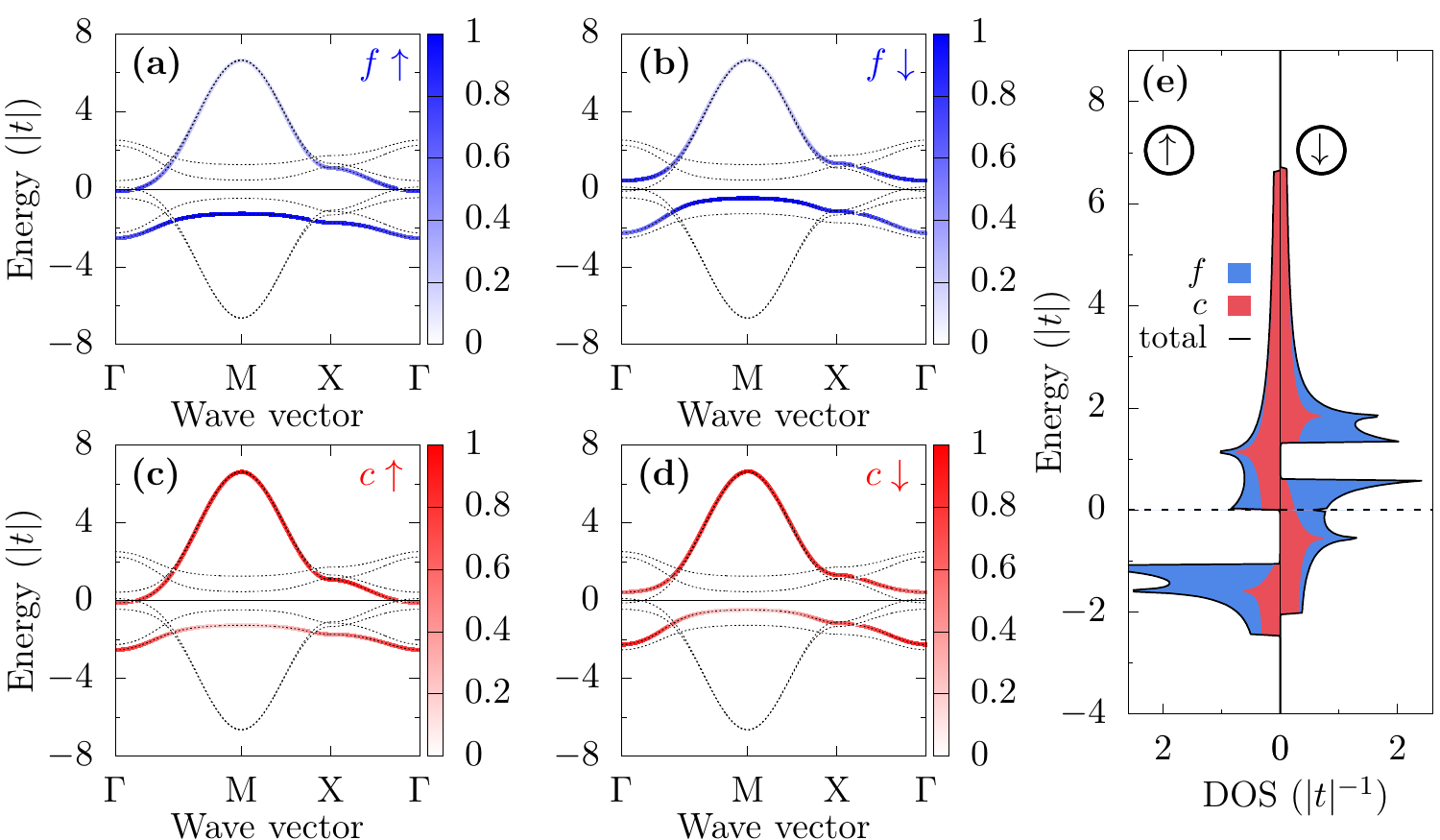}
\caption{(a)-(d) Band structure for $h/|t| = 0.002$ and slightly larger hybridization $V/t = 1.262$ in the FM1+$A_1$ phase (the remaining parameters are same as in Fig.~\ref{fig:band_structure_U35_J11_FM2}). Eigenenergies are represented by dotted lines, partial contributions from $f$-electrons are marked in blue, whereas those from $c$-electrons in red. Color intensity represents the spectral weight. (e) Orbital- and spin-resolved density of states.}
	\label{fig:band_structure_U35_J11_FM1}
\end{figure*}

FM and SC  phase transitions have a substantial impact on the electronic (renormalized-band) structure. Particularly interesting is the situation near the boundary between FM2+$A_2$ and FM1+$A_1$ states. To elucidate the changes on both sides of the transition, in Figs.~\ref{fig:band_structure_U35_J11_FM2} and \ref{fig:band_structure_U35_J11_FM1} we have plotted an exemplary structure along the high-symmetry lines just below ($V/t = 1.26$) and just above that value ($V/t=1.262$)  for $h=2\cdot 10^{-3}|t|$. The slightly different magnitudes of $V$ have been selected to visualize the situation on both sides of the discontinuous FM1$\rightarrow$FM2 transition. The spin subbands with the dominant $f$ character are marked in blue. The spin splitting is induced mainly by the Hund's rule and on-site repulsion $U$ (the effects of applied field and pairing are of minor importance). Remarkably, the $c$ electrons (marked in red in the lower panel) exhibit also a comparable spin splitting. This effect is caused by the circumstances that the $c$ electrons are hybridized with their $f$ electron partners and, therefore, the Hund's rule interaction is transferred from $f$- to $c$-system. Note that the occupancy of each of the $f$ orbitals is $n^f/2 =1 \pm \delta$, with $\delta\ll 1$ [cf. Fig \ref{fig:occupancy}(b)], where the small portion $\delta$ comes from the upper spin subband which crosses the Fermi (zero-energy level) near $\Gamma$ point. This is explicitly visualized on the density of states (right panel), where the $f\downarrow$ subband barely touches the Fermi energy and the majority spin subband is practically filled. To a good accuracy, the system is thus a \textit{half metal} with the predominant spin-minority carriers at the Fermi level. This is the reason why the ${A_2}$ phase is stable then and with the amplitudes $\Delta_{\downarrow\downarrow} \gg \Delta_{\uparrow\uparrow }$. The situation turns into an extreme case, with only $\Delta_{\downarrow\downarrow}\neq 0$ in the FM1+${A_1}$ phase. The latter result rationalizes nicely the related observation in $H=0$ situation.\cite{Kadzielawa-MajorPhysRevB2018} Note, however, that the exact half-metallic behavior, obtained in the present model, might be obscured in the real material by other bands that are weakly coupled to the considered $f$-$c$ subsystem.

\section{Outlook}

\subsection{Effect of spin fluctuations (tentative)}

Our present approach, based on first nontrivial order (SGA) of treating the interelectronic correlations on local scale, cannot explain enhanced residual linear specific heat appearing at temperatures well below $T_c$ in $\mathrm{UGe_2}$,\cite{TateiwaPhysRevB2004} as well as the strong effective mass enhancement at the FM1$\rightarrow$FM2 transition there.\cite{TerashimaPhysRevEltt2001,AokiJPhysSocJapan2019} Additionally, NQR relaxation with an anomalous temperature dependence is observed at the FM to PM transition at low temperature.\cite{ManagoPhysRevB2019} All these features may be explained qualitatively in terms of FM spin fluctuations starting from our renormalized mean-field picture. Whereas overall features of a transition from non-unitary to unitary SC are well reproduced by our phase diagram (also for $\mathrm{UTe_2}$, cf. Ref.~[\!\!\citenum{RanArXiV2018}]), the long-wavelength fluctuations should be included, particularly for low-moment bearing systems $\mathrm{UCoGe}$ and $\mathrm{UIr}$.

A general way to extend our work is as follows. We start from the effective Hamiltonian \eqref{eq:hamiltonian}, but with renormalized microscopic parameters $U \lambda_{\uparrow\downarrow}^2$ and $J g_{2\sigma}$ [cf. Eq.~\eqref{eq:energy_functional}], and proceed with the Hubbard-Stratonovich transformation, as outlined in Appendix~\ref{appendix:fluctuations} for the case of FM state. To incorporate fluctuations of the SC order parameter, one should include also the bilinear representation of the spin part $\sim J g_{2\sigma}$, derived in Ref.~[\!\!\citenum{SpalekPhysRevB2001}].  However, a quantitative implementation of this program is quite cumbersome, as it requires computation of renormalized coupling constants at each stage of the analysis, before and after including the fluctuations in each order. Nonetheless, we believe that such a solution is possible to tackle,  as the renormalized coupling parameters are reduced in the process already at the level of SGA. It is tempting to suggest that the effective picture should be not far away from that based on $1/N$ expansion with effective parameters $U$ and $J$ (cf.~Appendix~\ref{appendix:fluctuations}) calculated self-consistently within SGA.\cite{TakahashiBook2013}
\subsection{Summary}

In the preceding paper, \cite{Kadzielawa-MajorPhysRevB2018} regarded here as Part I, we have constructed a fairly complete zero-magnetic-field phase diagram composed of spin-triplet paired states  coexisting with the ferromagnetic FM1 and FM2 phases. The ${A_2}$ and $A$ SC states appear in the field absence only with very small amplitudes. In the present work we have shown that  the applied magnetic field allows for fine tuning of those phases and is likely to make them observable. In this manner, one can detect the states analogous to those seen clearly only for the superfluid $\mathrm{^3He}$.\cite{Vollhardt1990} However, in contrast to $\mathrm{^3He}$, here the pairing is of $s$-wave character, i.e., with intraatomic spin-triplet and the orbital singlet to make the wave function of the local pairs antisymmetric. It should be emphasized that this picture is applied here for  moderately correlated systems, in which the pairing is induced by the Hund's rule combined with direct short-range Coulomb interaction. In the strong-correlation limit, this type of pairing would have intersite (real-space) character with either spin-triplet or spin-singlet nature, depending on the band filling.\cite{KlejnbergJPhysCondensMatter1999,KlejnbergPhysRevB2000}

The principal result of this and the preceding\cite{Kadzielawa-MajorPhysRevB2018} work is to describe, within a single (orbitally degenerate) Anderson lattice model, coexistent FM and spin-triplet SC phases within a consistent picture. In this way, we extend the well established approaches to correlated normal and magnetic systems\cite{SpalekPhysRevLett1987} to include the SC states coexisting with them and within a single mechanism. It must be emphasized that such a renormalized mean field theory may be also generalized to a more involved systematic form of diagrammatic expansion, DE-GWF.\cite{BunemannPhyRevB1998,SpalekJPhysCondensMatter2013} However, such an approach becomes quite involved in the multi-orbital situation, particularly with multiple coexisting phases.\cite{ZegrodnikNewJPhys2013,ZegrodnikNewJPhys2014} Inclusion of higher-order correlations introduces then an additional admixture of intersite correlations to the present local pairing. This should be an objective of a separate study.

At the end, we should mention that the present model neglects spin-orbit coupling and magnetocrystalline anisotropy in the uranium compounds addressed above.\cite{ShickPhysRevLett2001,ShickPhysRevB2004} From the fact that the overall phase diagram and the coexistent phases are reproduced correctly, we draw the conclusion that the orbital moment may be frozen (we consider only spin-aligned phases) and that the anisotropic character of the system in enforced naturally by the presence of the long-range FM order along the easy axis. Obviously, this may not be that simple if we would like to discuss the situation in the field by changing its orientation.

\acknowledgments{This work was supported by the Grant OPUS No. UMO-2018/29/B/ST3/02646 from Narodowe Centrum Nauki (NCN).}

\appendix

\section{Statistically Consistent Gutzwiller Approximation (SGA): Simplified vs. full forms}
\label{appendix:SGA}

In the preceding paper\cite{Kadzielawa-MajorPhysRevB2018} (cf. Appendix~A there) we have discussed in detail the SGA approximation. Here we provide a more formal background. First, the multi-band trial function for the ground state is selected in the form
\begin{equation}
\ket{\Psi_G} = \prod_i \hat{P}_i \ket{\Psi_0},
\label{eq:psi_g}
\end{equation}
where $\ket{\Psi_0}$ is  an antisymmetrized product (Slater determinant) of single-particle wave functions, in general describing non-correlated broken symmetry state, for which the Wick's theorem holds. Operator $\hat{P}_i$ is the so called Gutzwiller correlator that changes the weights of various many-body configurations in the variational wave function $\ket{\Psi_G}$. The general form of $\hat{P}_i$ is
\begin{equation}
\hat{P}_i  = \sum_{II'} \lambda_{i\,II'} \ketbra{I,i}{I',i},
\end{equation}
where the states $\{\ket{I,i}\}_I$ span the local Fock space of the correlated orbitals at site $i$ and the variational variables $\lambda_{iII^\prime}$ form a matrix, here taken in the real-valued and symmetric form. Those correlated local spin-orbital states can be represented as 
\begin{equation}
\ket{I,i} = \prod^{<}_{\alpha \in I} \hat{f}^{\dagger}_{i\alpha} \ket{0,i},
\label{eq:local_state}
\end{equation}
where $\alpha = (l, \sigma)$ labels combined spin-orbital indices and the symbol `$<$' indicates a specified selected ascending order of the creation operators. Likewise,
\begin{equation}
\bra{I',i} = \prod^{>}_{\alpha \in I'}  \bra{0,i} \hat{f}_{i\alpha}
\label{eq:local_state_2}
\end{equation}
contains the annihilation operators in the descending order so that 
\begin{equation}
\ketbra{I,i}{I',i} = \prod^{<}_{\alpha \in I} \hat{f}^{\dagger}_{i\alpha} \prod^{>}_{\beta \in I'} \hat{f}^{}_{i\beta} \prod_{\gamma \notin I \cup I'} (1 - \hat{n}^f_{i\gamma}).
\label{eq:transition_operator}
\end{equation}

The basic task is to compute the ground state energy. For that purpose, one needs to evaluate the averages of the form
\begin{equation}
\expval**{\hat{O}_i}{\Psi_G} = \frac{\expval**{\qty\Big(\prod_j \hat{P}_j) \hat{O}_i \qty\Big(\prod_j \hat{P}_j) }{\Psi_0}}{ \expval**{\qty\Big(\prod_j \hat{P}_j) \qty\Big(\prod_j \hat{P}_j) }{\Psi_0}}
\label{eq:expval_local}.
\end{equation}
The products of local correlators can be now rearranged by using the fact that $\hat{P}_i$ and $\hat{P}_j$ commute for $i\neq j$. In effect,
\begin{equation}
\expval{\hat{O}_i} = \frac{\Big<\qty\Big(\prod_{j\neq i} \hat{P}_j^{\,2}) \hat{P}_i \hat{O}_i \hat{P}_i\Big>_0}{\Big< \prod_{j} \hat{P}_j^{\,2} \Big>_0}
\label{eq:expval_local_2}, 
\end{equation}
where the averages with the subscript ''0'' are taken in the uncorrelated state, so that when applied the Wick theorem to the averages in the uncorrelated $\langle\ldots \rangle_0$ representation, we obtain 

\begin{align}
\Big< \prod_{j} \hat{P}_j^{\,2} \Big>_0 \expval{\hat{O}_i} &= 
\Big< \prod_{j\neq i} \hat{P}_j^{\,2}\Big>_0 \expval{\hat{P}_i \hat{O}_i \hat{P}_i }_0
\nonumber
\\ 
& \contraction{
	+ \sum_{\substack{\text{\tiny all pairs} \\ \text{\tiny of n. n.} \\\text{\tiny contractions}}} }
	{\Big<\prod_{j\neq i} \hat{P}_j^{\,2}\Big>_0}
	{}
	{\Big<\hat{P}_i \hat{O}_i \hat{P}_i\Big>_0}
\contraction[2ex]{
		+ \sum_{\substack{\text{\tiny all pairs} \\ \text{\tiny of n. n.} \\\text{\tiny contractions}}} \Big<}
	{\prod_{j\neq i} }
	{\hat{P}_j^{\,2}\Big>_0 \Big<\hat{P}_i }
	{\hat{O}_i \hat{P}_i\Big>_0}
+ \sum_{\substack{\text{\tiny all pairs} \\ \text{\tiny of n. n.} \\\text{\tiny contractions}}}
\Big<\prod_{j\neq i} \hat{P}_j^{\,2}\Big>_0 \Big<\hat{P}_i \hat{O}_i \hat{P}_i\Big>_0 + \ldots,
\end{align}
where the symbol
\begin{align}
\contraction[1.8ex]{
\sum_{\substack{\text{\tiny all pairs} \\ \text{\tiny of n. n.} \\\text{\tiny contractions}}} }
{\Big< \hat{\mathcal{A}} }
{\Big>_0}
{ \Big< \hat{\mathcal{B}}\Big>_0}
\contraction{
\sum_{\substack{\text{\tiny all pairs} \\ \text{\tiny of n. n.} \\\text{\tiny contractions}}} }
{\Big< \hat{\mathcal{A}} \Big>_0}
{}
{\Big< \hat{\mathcal{B}}}
\sum_{\substack{\text{\tiny all pairs} \\ \text{\tiny of n. n.} \\\text{\tiny contractions}}}
\Big<\hat{\mathcal{A}}\Big>_0 \Big<\hat{\mathcal{B}}\Big>_0
\end{align}
represents all the nonzero pair contractions selected for a given broken-symmetry state. A detailed procedure is quite cumbersome and will not be detailed here.\cite{KubiczekMasterThesis,BunemannPhyRevB1998,SpalekJPhysCondensMatter2013} 

Under the so-called Gutzwiller conditions \cite{BunemannPhyRevB1998,SpalekJPhysCondensMatter2013,KubiczekMasterThesis}

\begin{align}
\Big< \hat{P}_i^{\,2}\Big>_0 &= 1, \\
\Big< \hat{P}_i^{\,2} \hat{f}^{\dagger}_{i\alpha} \hat{f}^{}_{i\beta} \Big>_0 &= \Big< \hat{f}^{\dagger}_{i\alpha} \hat{f}^{}_{i\beta} \Big>_0,
\end{align}

\noindent
and for large site-coordination number, a straightforward general formula for the expectation values of local operators is obtained

\begin{equation}
\expval**{\hat{O}_i}{\Psi_G} = \expval{\hat{P}_i \hat{O}_i \hat{P}_i }_0,
\end{equation}

\noindent
which can be used to evaluate $\expval**{\mathcal{H}}{\Psi_G}$ (note that all the $f$-dependent terms are local).

In effect, we obtain Landau-type functional $L$ which, at $T=0$, is composed of $\expval{\mathcal{H}}_{G}$ and incorporates the condition for the chemical potential, the enforced normalization $\braket{\Psi_{G}}=1$, as well as the requirement of having the same number of particles in the initial ($\ket{\Psi_0}$) and correlated ($\ket{\Psi_G}$) states (before and after
the projection with $\hat{P}_G$), namely

\begin{align}
L & \equiv \expval{\hat{H}}_{G}  - \mu \sum_i \qty\bigg(\sum_{\alpha}\expval{\hat{n}^f_{i\alpha}} + \sum_{\beta}\Big< \hat{n}^c_{i\beta}\Big> - n^\mathrm{tot}) \nonumber \\  & + \sum_{i} \eta_i \qty\bigg(\expval{\hat{P}_i^{\,2}}_0 - 1)  \nonumber \\  &
 +\sum_{i\alpha\beta} \eta_{i\alpha\beta}  \qty\bigg(\expval{\hat{P}_i^{\,2}\hat{f}^{\dagger}_{i\alpha} \hat{f}^{}_{i\beta} }_0 - \expval{\hat{f}^{\dagger}_{i\alpha} \hat{f}^{}_{i\beta} }_0). 
\label{eq:lagrangian}
\end{align}

\noindent
The functional $L$ needs to be optimized with respect to all $\lambda$ and $\eta$ parameters, representing additional constraints \cite{KubiczekMasterThesis} in the SGA approximation, as well as $\mu$ and $\ket{\Psi_0}$. Minimization with respect to $\ket{\Psi_0}$ leads to an effective (renormalized) quasiparticle Hamiltonian in an applied magnetic field $h$ which, in the component Nambu representation [cf. Eq.~\eqref{eq:effective_hamiltonian}], can be recast to the following form

\begin{equation}
  \mathcal{H}_{\mathrm{eff}} = \sum_{\mathbf{k} \sigma}
  \Psi^\dagger_{\mathbf{k}\sigma}
\begin{pmatrix}
\epsilon_{\mathbf{k}\sigma}& 0 & \tilde{V}_{\sigma} & \tilde{\Delta}_{fc\sigma} \\
0 & -\epsilon_{\mathbf{k}\sigma} & \tilde{\Delta}_{fc\sigma} & -\tilde{V}_{\sigma} \\
\tilde{V}_{\sigma} & \tilde{\Delta}_{fc\sigma} & \tilde{\epsilon}_{f\sigma} & \tilde{\Delta}_{f\sigma} \\
\tilde{\Delta}_{fc\sigma} & -\tilde{V}_{\sigma} & \tilde{\Delta}_{f\sigma} & -\tilde{\epsilon}_{f\sigma}
\end{pmatrix}
\Psi_{\mathbf{k}\sigma} + E_0,
\label{eq:H_eff_Nambu}
\end{equation}

\noindent
with the renormalized parameters defined as

\begin{align}
  \tilde{\epsilon}_{f\sigma} &\equiv \frac{1}{2} \pdv{L}{n^0_{f\sigma}}, \label{eq:def_epsilon_f}\\
  \tilde{V}_{\sigma} &\equiv \frac{1}{4} \pdv{L}{v^0_{\sigma}}, \\
\tilde{\Delta}_{f\sigma} &\equiv \frac{1}{2} \pdv{L}{A^0_{f\sigma}}, \label{eq:def_Delta_ff}\\
\tilde{\Delta}_{fc\sigma} &\equiv \frac{1}{4} \pdv{L}{A^0_{fc\sigma}} \label{eq:def_Delta_fc},
\end{align}

\noindent
and  $\epsilon_{\mathbf{k}\sigma}$ given by Eq.~\eqref{eq:zeeman_split_dispersion}. The bare parameters read

\begin{align}
 n^0_{f\sigma}  &\equiv \expval{\hat{f}^{\dagger}_{il\sigma} \hat{f}^{}_{il\sigma}}_0, \label{eq:n0}\\
  n^0_{c\sigma}  &\equiv \expval{\hat{c}^{\dagger}_{il\sigma} \hat{c}^{}_{il\sigma}}_0 \label{eq:nc},\\
 v^0_{\sigma}  &\equiv \expval{\hat{f}^{\dagger}_{il\sigma} \hat{c}^{}_{il\sigma}}_0,\\
A^0_{f\sigma}  &\equiv \expval{\hat{f}^{\dagger}_{i1\sigma} \hat{f}^{\dagger}_{i2\sigma}}_0, \\
A^0_{fc\sigma}  &\equiv \expval{\hat{f}^{\dagger}_{i1\sigma} \hat{c}^{\dagger}_{i2\sigma}}_0 = \expval{\hat{c}^{\dagger}_{i1\sigma} \hat{f}^{\dagger}_{i2\sigma}}_0. \label{eq:A0}
\end{align}
Note that the averages (\ref{eq:n0})-(\ref{eq:A0}) define the uncorrelated broken-symmetry state, whereas Eqs.~\eqref{eq:def_epsilon_f}-\eqref{eq:def_Delta_fc}  define the physical state. Also, the Hamiltonian (\ref{eq:H_eff_Nambu}) is self-consistent in which the quantities defining the physical state are $\mu, \tilde{\Delta}_{f\sigma}, \tilde{\Delta}_{fc\sigma}, \tilde{V}_{\sigma}, \tilde{\epsilon}_{f\sigma}$, and the band dispersion relation of $\epsilon_{\bf k \sigma}$ for bare $c$ electrons. They are determined from a system of five self-consistent equations. Note also that in the effective Hamiltonian~\eqref{eq:effective_hamiltonian} the anomalous averages $\tilde{\Delta}_{fc\sigma}$ are set to zero, which means that the direct hybrid ($c$-$f$) pairing is regarded as negligible. This is not the case for the singlet-paired systems.\cite{WysokinskiPhysRevB2016,KarbowskiPhysRevB1994,HowczakPhysStatSolidiB2013}

\begin{figure}
  \includegraphics[width = 0.99\columnwidth]{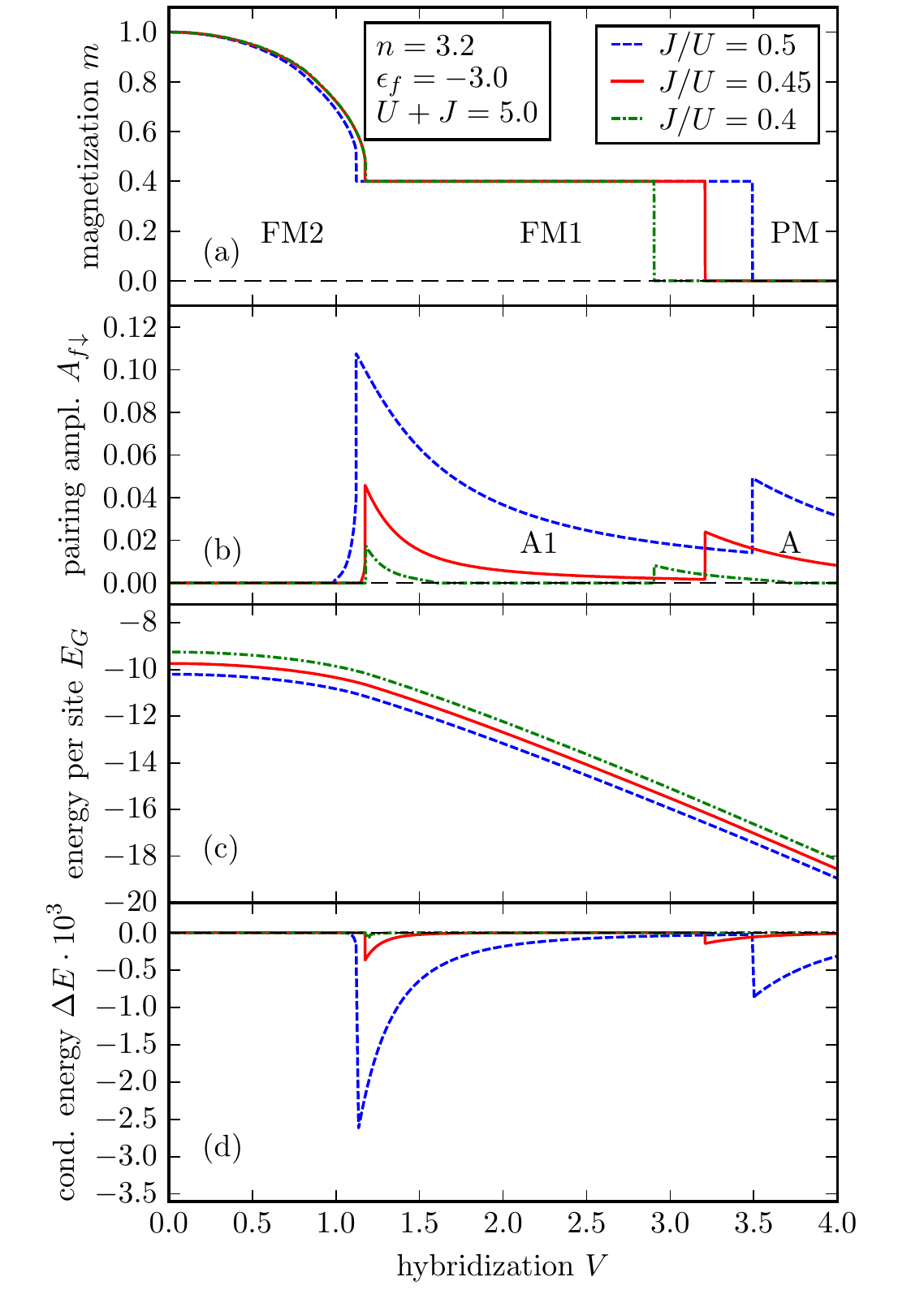}
  \caption{Exemplary phase diagram obtained with the multi-orbital correlator in the $f$-electron sector.  (a) The total magnetization $m$, (b) pairing amplitude $A_{f\downarrow}$, (c) ground state energy $E_{G}$ per lattice site, (d) SC condensation energy $\Delta E$, all as a function of hybridization for $n=3.2$, $\epsilon_{f}=-3|t|$, $U+J=5|t|$, square lattice density of $c$ states: $t < 0$, $t'=0.25|t|$ and for three rations $J/U=0.5, 0.45, 0.4$. Note that ${A_1}$ phase is characterized by $A_{f\uparrow}=0$, whereas in A phase we have $A_{f\uparrow}=A_{f\downarrow}$. The condensation energy for $J/U=0.4$ is so low that it is hardly visible on the scale. Note that despite seemingly discontinuous behavior, $\Delta E$ does not exhibit jumps across the joint metamagnetic and metasuperconducting transitions, but it varies extremely rapidly in the narrow parameter range. For the zero-field case this has been detailed in Appendix~D of Ref.~[\!\!\citenum{KadzielawaMajorPhDThesis}].
  }
  \label{fig:hybridization}
\end{figure}

In Fig.~\ref{fig:hybridization}(a)-(d) we display the selected properties of SC state on the basis of full solution of the self-consistent equations obtained with the help of Hamiltonian (\ref{eq:H_eff_Nambu}), for the three selected values of the Hund's rule exchange integral $J$. Namely, in (a) we display the total magnetic moment $m^\mathrm{tot}$. Panel (b) shows the dominant (spin-down) pairing amplitude in FM1+${A_1}$ and $\mathrm{PM}$+$A$ phases. In panel (c) we draw the ground-state  energy, whereas in (d) we plot the condensation energy (the energy difference between the SC state and that corresponding to the appropriate pure FM phase). All these characteristics are quantitatively similar to those obtained earlier within the simplified picture with $\Delta_{fc\sigma}\equiv 0$. From that we draw the conclusion that the hybrid pairing component has a negligible effect on SC. Also, the component $\Delta_0$ of the pairing amplitude of $f$-electrons (i.e., the one with zero $z$ spin-component of the pair) is suppressed in this system with relatively large $U$. Hence, the simplified solution detailed in Appendix~A of Ref.~[\!\!\citenum{Kadzielawa-MajorPhysRevB2018}] represents, to a good accuracy, the full solution. The same type of picture is used throughout  the present paper for the case of nonzero applied field. 

\section{Incorporation of quantum spin fluctuations in an orbitally degenerate system: An outline}
\label{appendix:fluctuations}

The atomic part of the Hamiltonian~\eqref{eq:hamiltonian} for $f$ electrons located on orbitals $l=1, 2, \ldots, d$, where $d$ is their degeneracy, can be rewritten in the form

\begin{align}
  \label{eq:eq_app_fluct:H_I}
  \mathcal{H}_I =& U \sum\limits_{il} \hat{n}^{f(l)}_{i\downarrow} \hat{n}^{f(l)}_{i\uparrow} + \frac{K}{2} \sum\limits_{\overset{ill'}{\sigma\sigma'}} {}^{'} \hat{n}^{f(l)}_{i\sigma} \hat{n}^{f(l')}_{i\sigma'} \nonumber\\& - J \sum\limits_{ill'} \hat{\mathbf{S}}^{f(l)}_{i} \hat{\mathbf{S}}^{f(l')}_{i},
\end{align}

\noindent
where $K=U' - J/2$ and the primed summation is performed over $l \neq l'$. Note that the interaction parameters $U$, $K$, and $J$ are taken as the same for each pair $(l, l^\prime)$ of orthogonalized orbitals. Therefore, we introduce next the global spin- and particle-number operators as

\begin{align}
  \label{eq:eq_app_fluct:global_operators}
  \hat{\mathbf{S}}^f_i \equiv \sum\limits_{l = 1}^d \hat{\mathbf{S}}^{f(l)}_{i}, \hspace{2em}   \hat{n}^f_i \equiv \sum\limits_\sigma \hat{n}^{f}_{i\sigma} \equiv \sum\limits_{l\sigma} \hat{n}^{f(l)}_{i\sigma}.
\end{align}

\noindent
By expressing the orbital-dependent operators in Eq.~\eqref{eq:eq_app_fluct:H_I} through their global correspondants,\cite{KlejnbergPhysRevB1998} up to a constant, one obtains

\begin{align}
  \label{eq:eq_app_fluct:H_I2}
  \mathcal{H}_I =& \frac{1}{2} K \sum\limits_{i} \left(\hat{n}^f_{i}\right)^2 - J \sum\limits_{i} \left(\hat{\mathbf{S}}^f_{i}\right)^2 + I \sum\limits_{il} \hat{n}^{f(l)}_{i\uparrow} \hat{n}^{f(l)}_{i\downarrow}
\end{align}

\noindent
with $I\equiv U - K - \frac{3}{2} J$. Assuming the standard relation for $d$ electrons $U' = U - 2J$ we obtain $K = U - \frac{5}{2} J$, $I = J/2$. We thus have decomposed the intraatomic interaction into the three parts: local charge, spin, and the Hubbard-type correlations, respectively. Now, noticing that the first two terms give contribution of the order of $d^2$, whereas the third one $\sim d$, and disregarding charge fluctuations, we have, to the first approximation,

\begin{align}
  \label{eq:eq_app_fluct:H_I_approx}
  \mathcal{H}_I = - \left(J + \frac{I}{3d}\right) \sum\limits_{i} \left(\hat{\mathbf{S}}^f_{i}\right)^2,
\end{align}

\noindent
i.e., the total local spin fluctuations provide the leading contribution. In the FM state one can take $\hat{\mathbf{S}}_i^f = \langle S_i^{fz}\rangle \hat{\mathbf{e}}_z + \hat{\mathbf{s}}_i$, where the static part of magnetization introduces a natural anisotropy axis for spin fluctuations expressed by $\hat{\mathbf{s}}_i = \hat{\mathbf{s}}_i(\tau)$, where $\tau$ is the imaginary time. To include the dynamic fluctuations one utilizes the Hubbard-Stratonovich transformation

\begin{align}
  \label{eq_app_fluct:HS_transformation}
  \exp(\hat{a}^2) = \int\limits_{-\infty}^{\infty} dx \exp(-\pi x^2 - 2\hat{a}x\sqrt{\pi})
\end{align}

\noindent
for each spin-operator component $\hat{S}^{f\alpha}(\tau)$. By including also the single-particle part $\hat{\mathcal{H}}_0$, we obtain the following expression for the system density matrix

\begin{widetext}
\begin{align}
  \label{eq_app_fluct:density_matrix}
  \rho = T\mathrm{e}^{-\beta\hat{\mathcal{H}}_0}\prod\limits_i \int \mathcal{D} \xi_i^\alpha(\tau) \exp(-\int \limits_0^1 d\tau (\xi_i^\alpha)^2 - \int \limits_0^1 d\tau 2i\sqrt{\pi\beta J} \xi_i^\alpha(\tau) A_i^\alpha(\tau) ),
\end{align}

\end{widetext}

\noindent
where $\mathcal{D} \xi_i^\alpha(\tau)$ denotes functional integration over each Gaussian random field $\xi_i^\alpha(\tau)$, $\beta \equiv (k_BT)^{-1}$, $\tau$ is in the units of $\beta$, and $A_i^\alpha(\tau) \equiv \hat{S}_i^{f\alpha}(\tau)$. One can see that this form is of the same type as that for the Hubbard with explicitly rotationally invariant interaction term

\begin{align}
  \label{eq_app_fluct:rotationally_inv_hubbard}
  U \hat{n}_{i\uparrow} \hat{n}_{i\downarrow} = \frac{1}{4} U (\hat{n}_{i\uparrow} + \hat{n}_{i\downarrow})^2 - \frac{1}{3}U \hat{\mathbf{S}}_i^2
\end{align} 

\noindent
and fluctuating field $\hat{S}_i^{\alpha}(\tau)$.\cite{EvensonJApplPhys1970} Therefore, the spin-fluctuation contribution can be calculated in the same manner as in the Hubbard model with the part $\langle S_i^z\rangle \neq 0$. However, in order to incorporate the fluctuations starting from the SGA (renormalized mean-field) solution, replacing the Hartree-Fock solution as a saddle-point approximation, our coupling constant must be also renormalized, $J \rightarrow J\lambda_J$ as contained when solving self-consistent equation for $\langle S^{fz}_i\rangle_0$. Implementation  of this program is quite involved, both analytically and numerically, so it should be analyzed in detail separately. In any case, the spin-fluctuation contribution will renormalize the SGA characteristics by \textit{not} just an additive  contribution. However, a further generalization of expression \eqref{eq:eq_app_fluct:H_I_approx} is required to include also the pairing fluctuations. This can be implemented in the following manner. We start from the binomial representation of the Hund's rule part which, for the simplest spin $S=1$ case ($l = 1, 2$), takes the form

\begin{align}
  \label{eq_app:binomail_representation}
  \hat{\mathbf{S}}_i^{f(l)}  \hat{\mathbf{S}}_i^{f(l')} + \frac{3}{4} \hat{n}_i^{f(l)} \hat{n}_i^{f(l')} = \sum\limits_{m=-1}^1 \hat{A}^{f\dagger}_{im}\hat{A}^{f}_{im},
\end{align}

\noindent
where the pairing amplitude components are defined as\cite{SpalekPhysRevB2001}

\begin{align}
  \label{eq_app:pairing_amplitudes_def}
  \begin{cases}
    \hat{A}^\dagger_{i1} \equiv \hat{f}^{(1)\dagger}_{i\uparrow} \hat{f}^{(2)\dagger}_{i\uparrow}, \\
    \hat{A}^\dagger_{i0} \equiv \frac{1}{\sqrt{2}} \left(\hat{f}^{(1)\dagger}_{i\uparrow} \hat{f}^{(2)\dagger}_{i\downarrow} + \hat{f}^{(1)\dagger}_{i\downarrow} \hat{f}^{(2)\dagger}_{i\uparrow}\right), \\
    \hat{A}^\dagger_{i-1} \equiv \hat{f}^{(1)\dagger}_{i\downarrow} \hat{f}^{(2)\dagger}_{i\downarrow}. \\
  \end{cases}
\end{align}

\noindent
This bilinear form can be transformed to the corresponding representation \eqref{eq_app_fluct:density_matrix} and will involve additional fluctuating fields $\{\eta^m_{i}(\tau)\}$ ($m=-1,0,+1$) which express three local components of the pairing $\Delta_{im}^f$. In general, one can decompose the Hund's rule term into two components, diagonal (magnetic moment) and off-diagonal (pairing gap) according to the prescription provided in Ref.~[\!\!\citenum{LindnerJPhysCondesMatter1991}].

\end{document}